 \documentstyle[12pt]{article}

\renewcommand{\a}{\alpha}
\renewcommand{\b}{\beta}
\newcommand{\g}{\gamma}

\renewcommand{\d}{\delta}
\newcommand{\D}{\Delta}

\newcommand{\h}{\eta}
\newcommand{\k}{\kappa}
\renewcommand{\l}{\lambda}
\newcommand{\m}{\mu}
\newcommand{\n}{\nu}
\newcommand{\p}{\phi}
\newcommand{\r}{\rho}
\newcommand{\s}{\sigma}
\renewcommand{\t}{\tau}
\newcommand{\th}{\theta}
\newcommand{\na}{\nabla}
\newcommand{\z}{\zeta}
\def\ssc{\scriptscriptstyle}
\def\I#1{I_{\ssc #1}}
\def\etal{{\it et al.}\ }
\def\ie{{\it i.e.,}\ }
\def\eg{{\it e.g.,}\ }
\def\ap{\alpha'}

\begin{document}

\begin{titlepage}
\begin{flushright}    {\small McGill/94-48 \\  hep-th/9411018 }\\
\end{flushright}
\vskip 5em

\begin{center}
{\bf \huge Gravity's Rainbow:} \\
{\bf \Large Limits for the applicability of the equivalence principle}
\vskip 3em

{\large Ren\'e Lafrance\footnote{lafrance@physics.mcgill.ca}
  and  Robert C. Myers\footnote{rcm@physics.mcgill.ca}}\\
 \vskip 1em
  Department of Physics, McGill University \\
	Montr\'eal, Qu\'ebec, Canada H3A 2T8

 \vskip 4em

\begin{abstract}
Limits for the applicability of the equivalence principle are considered
in the context of low-energy effective field theories. In particular,
we find a class of higher-derivative
interactions for the gravitational and electromagnetic
fields which produce dispersive photon propagation.
The latter is illustrated by calculating the energy-dependent
contribution to the deflection of light rays.
\end{abstract}
\end{center}

\vskip 3em
PACS numbers: 04.40.Nr, 11.25.Mj
\end{titlepage}

\section{Introduction}
The deflection of light by a gravitational field was one of the most
striking predictions of general relativity. The observation of this
effect was also one of the earliest experimental verifications of
Einstein's theory\cite{edd}. The topic has presently matured to the
stage where gravitational lensing is applied by astronomers for
a variety of purposes\cite{lense}, and in particular, it is providing
an exciting new probe of the dark matter\cite{macho}.

By the equivalence principle, photons fall
freely along light-like geodesics in curved spacetime, and
all photons are deflected in a gravitational potential by the same angle
independently of their energy or their polarization.
Things change when one goes beyond the standard
minimal coupling of the electromagnetic
and gravitational fields. For example in quantum electrodynamics,
virtual electron loops will induce curvature couplings in the
effective action of low-frequency photons\cite{Berends,loop2}.
In such a case, photon propagation can be influenced by
tidal effects coming from local spacetime curvature.
Using the one-loop effective action
for QED, Drummonds and Hathrell \cite{loop2}
studied the propagation of photons in Schwarzschild,
Robertson-Walker and gravitational-wave spacetimes.
Apart from the Robertson-Walker background where the curvature is
spatially isotropic, they
found that the propagation of photons was polarization-dependent
(gravitational birefringence). In particular, they considered
light rays following transverse orbits in the Schwarzschild background, and
 found a polarization-dependent deflection angle.
Similarly, gravitational birefringence was
also found to occur in Reissner-Norstr\"om spacetimes \cite{Daniels}.

In the context of string theory, Mende \cite{Mende} also
observed limits for the applicability of the equivalence
principle. Since strings are extended objects, their propagation
also feels tidal effects arising from the curvature of spacetime.
Mende argued that this would imply
energy-dependent deflection of light. Specifically, for photons
following transverse orbits passing a spherically symmetric
mass, the deflection angle would have a contribution
proportional to the square of the photon energy.
Further Mende argued that such a result would be a clear signature
of string theory. We will show that such energy-dependent light
scattering can also be produced within the context of
a low-energy effective action for the electromagnetic field
in curved spacetime \cite{foot1}.

The paper is organized as follows: In section 2, we establish
the framework of our calculations. We review the geometric optics
approximation,
and illustrate how higher-derivative interactions can modify
the photon trajectories. The form of interactions which can
lead to photon propagation with dispersion is also established.
In section 3, we consider various eight-derivative interactions, and
show that they produce dispersive effects, such
as energy-dependent deflection of light.
Finally, section 4 presents a discussion of our results.
Throughout the paper, we employ the conventions of ref. \cite{MTW},
and we use units where $\hbar=c=G=1$.

\section{Geometric Optics Approximation}\label{6dsect}

Here, we will give a brief review of the geometric optics approximation
--- for more details, see ref.~\cite{MTW} and the references therein.
We begin with the Maxwell
action,
\begin{equation}
\I0 = -\frac{1}{4}\int d^4x \, \sqrt{-g} F_{\m\n}F^{\m\n}
\label{max}
\end{equation}
where $F_{\m\n}=\partial_\m A_\n-\partial_\n A_\m$. We impose the
Lorentz gauge condition
\begin{equation}
\nabla^\m A_\m=0\ ,
\label{lor}
\end{equation}
in which case the equation of motion may be written as
\begin{equation}
\nabla^\m F_{\m\n}=\nabla^2 A_\n-R_\n{}^\m A_\m=0\ \ .
\label{eom0}
\end{equation}
In the geometric optics approximation, one assumes that
the electromagnetic waves have a wavelength
which is much smaller than the radius of curvature of the
background geometry, or the scale for variations of
the amplitude of the wave front. The gauge potential is written as
\begin{equation}
A_\m= a_\m\ \exp[i\th]
\label{ans}
\end{equation}
where $a_\m$ is a slowly varying amplitude, and $\th$, a
rapidly varying phase. (It is understood that one takes the real part of the
right hand side of eq.~(\ref{ans})). The wave vector $k_\m=\nabla_\m\th$ is
normal to surfaces of constant phase, and light rays are defined by
${dx^\m\over ds}=k^\mu$. When this ansatz (\ref{ans}) is inserted
into the equation of motion and the gauge condition, the leading
terms come from the derivatives of the phase. Thus eq.~(\ref{eom0}) yields
\begin{equation}
k^\m k_\m=0
\label{lead1}
\end{equation}
while eq.~(\ref{lor}) produces
\begin{equation}
k^\m a_\m=0
\label{lead2}
\end{equation}
at leading order
--- \ie light rays are null geodesics, and the polarization is
orthogonal to the wave vector. At next order, eq.\ (\ref{eom0}) also yields
a propagation equation for the amplitude, $k^\m\nabla_\m a_\n+{1\over2}
(\nabla_\m k^\m) a_\n=0$. One can continue with a systematic
expansion of post-geometric-optics corrections, which would be
necessary to realize the full wave-like character of the
solutions --- \eg diffraction or interference. For our purposes,
though, we restrict our attention to the leading-order geometric-optics
equation. It is also assumed above
and in the following that the perturbation of the background metric
by the electromagnetic waves is negligible.

For our purposes, an important feature of the above leading-order equations
is that they are invariant under a constant scaling of the wave
vector, $k_\m\rightarrow \a k_\m$. Thus to this order, all solutions
will behave identically independent of the frequency of the photons.
Thus  physical effects, such as the bending of light rays, are
independent of the photon energy, $E$. Note that this scale invariance
does not hold for the post-geometric-optics corrections, which do
then yield frequency or energy dependent results. In the problem
of light ray deflection, these corrections, which produce
contributions proportional to $1/E$, signify diffractive (\ie
wave-like) effects. In contrast,
the effects, which we determine in the following section, are
proportional to $E^2$ and modify the light rays in the leading-order
approximation.

Now we will consider modifying the Maxwell action by adding
higher-derivative interactions, as may occur in an effective
field theory. Assuming that the former are quadratic in the
electromagnetic field strength,
they will generically modify the light-cone condition (\ref{lead1})
within the geometric optics approximation. Throughout the following, we
still impose the Lorentz gauge condition (\ref{lor}),
and so eq.~(\ref{lead2}) remains unchanged.
To illustrate the modifications produced by higher-derivative interactions,
consider
\begin{equation}
\I1=-{\a\l_e^2\over360\pi}\int d^4x \, \sqrt{-g}
\ R_{\m\n\s\t}  F^{\m\n} F^{\s\t}
\label{first}
\end{equation}
which arises as an interaction in the one-loop effective action
for QED\cite{Berends,loop2}. Here, $\l_e$ is the Compton
wavelength of the electron, and $\a$ is the fine structure
constant. There are other terms involving
$R_{\m\n}$ and $R$ , but they will be unimportant when considering
photon propagation in a background spacetime which is a solution of
the vacuum Einstein
equations. Combining eq.'s (\ref{max}) and (\ref{first}), the equation of
motion becomes
\begin{equation}
\nabla^\m F_{\m\n}+{\a\l_e^2\over90\pi}
(R_{\m\n\s\t}\nabla^\m +2\nabla_\s R_{\t\n})F^{\s\t}=0
\label{eom1}
\end{equation}
where the Bianchi identity for the Riemann tensor has been
applied in the last term. Inserting the geometric optics
ansatz (\ref{ans}) and applying the Lorentz gauge condition, we
find
\begin{equation}
k^2 a_\n + {\a\l_e^2\over 45\pi}R_{\m \n\s\t}k^\m k^\s a^\t=0
\label{new1}
\end{equation}
as the new leading order equation (where $k^2=k^\m k_\m$).
While the light-cone has been
modified away from $k^2=0$, the new equation is still invariant
when the wave-vector is scaled. Thus the resulting photon
trajectories are still frequency independent. The coupling
to the Riemann tensor in eq.~(\ref{new1}) does break local
Lorentz invariance, and this equation has been found to
produce polarization-dependent light propagation\cite{loop2}.

To have dispersive results then, the leading-order equation cannot
be invariant under scaling of $k_\mu$. Thus one must consider interactions
with more derivatives, and in particular, the interactions must
contribute to the electromagnetic equations of motion with more
derivatives of the gauge potential. With this in mind, a natural
extension of eq.~(\ref{first}) is then
\begin{equation}
\I2=-{\b\l^4\over4}\int d^4x \, \sqrt{-g}
\ R_{\m\n\s\t} \na_\r F^{\m\n} \na^\r F^{\s\t}
\label{second}
\end{equation}
where $\b$ is a dimensionless coupling constant, and $\l$ is the
(length) scale associated with the effective interaction. Combined with
the Maxwell action (\ref{max}), this new term leads to an
equation of motion
\begin{equation}
     \na_\m F^{\m\n} - \b\l^4 \na_\m \na_\r \left( R^{\m\n\s\t}
\na^{\r}F_{\s\t} \right) =0  \label{eom2}
\end{equation}
and in the leading order of the geometric optics
approximation, one finds
\begin{equation}
     k^2 \left( a_\n + 2 \b \l^4 R_{\m\n\s\t} k^\m k^\s a^\t\right) =0\ \ .
\label{new2}
\end{equation}
We will assume that the effective action is constructed {\it perturbatively}
in the coupling $\beta$. Within in such a framework,
even though eq.~(\ref{new2}) is not invariant under scaling of $k_\m$,
the light-cone condition remains $k^2=0$. The
second factor in brackets would define spurious characteristics which
are nonperturbative in $\b$. Alternatively, one may say that
perturbatively we wish to calculate to modification of
eq.~(\ref{lead1}) at order $\beta$,
and so expect to find $k^2=O(\b)$ in general.
Substituting the latter into the term proportional to $\b$
in eq.~(\ref{new2}), we
in fact have $k^2=O(\b^2)$ and so there is no perturbation of
the light-cone to the order which we are calculating.

There are other six-derivative interactions similar to $I_2$ where
the indices are contracted in  different ways, but in
the equations of motion, the
higher-derivative terms are proportional to $k^2$ again,
or to the gauge condition $k\cdot a$ which vanishes.
Thus we found that there are no six-derivative
interactions which will produce a dispersive light-cone condition.
To obtain an energy
dependent result, one needs to consider interactions with
both more derivatives and more background curvatures
in order to avoid the above contractions.

\section{Dispersive interactions}\label{sect8d}

  From the last section then, we have learned that in
order to produce a dispersive modification of the light-cone
condition, we need an interaction which is quadratic in
the field strength, has four derivatives of the gauge potential,
and has more than one background curvature or  derivatives of the
background curvature.
 In this section following these criteria,
we construct a number of eight-derivative interactions, and show
that they lead to energy-dependent photon propagation.
We begin with a simple extension
of eq.~(\ref{second}), where a second curvature tensor is introduced.
\begin{equation}
\I3=-\frac{\b\l^6}{4} \int d^4x\,\sqrt{-g}\ R^{\k\a\b\g} R_\k{}^{\r\m\n}
 \na_\a F_{\b\g} \na_\r F_{\m\n}
\label{third}
\end{equation}
where $\b$ and $\l$ are the coupling and scale, as above.
The equation of motion for the electromagnetic field becomes
\begin{equation}
\na_\m F^{\m\n} -\b \l^6\, \na_\m \na_\r \left( R^{\k\a\b\g}
R_\k{}^{\r\m\n} \na_\a F_{\b\g} \right) =0\ \ .
\label{eom3}
\end{equation}
This (along with the gauge constraint (\ref{lead2})) yield
in the geometric optics approximation,
\begin{equation}
  k^2 a^\n +2 \b \l^6
R^{\k\a\b\g} R_\k{}^{\r\m\n} k_\a k_\b k_\r k_\m a_\g =0   \ .
\label{new3}
\end{equation}
As desired, these equations are
not invariant under scaling of the wave vector,
and the higher order term is
not proportional to the original light-cone condition, $k^2$.
Hence these equations produce dispersion, and
should lead to energy-dependent light
deflection in a gravitational potential.

To explicitly display such dispersive light scattering,
we now turn to the specific background of the Schwarzschild metric in
standard coordinates \cite{MTW}
\begin{equation}
ds^2 = -\left( 1-\frac{2M}{r}\right) \, dt^2 + \left(
1-\frac{2M}{r}\right)^{-1} dr^2 +r^2 \left( d\theta^2 +\sin^2 \theta
\, d\p^2 \right)
\end{equation}
We introduce the vierbein $e^a{}_\m$ ($a=0,1,2,3$)
\[
   e^a{}_\m = {\rm diag} \left( U , 1/U ,
		r,r \sin \th \right)
\]
satisfying $g_{\m\n}=\eta_{ab}\,e^a{}_\m e^b{}_\n$
where $U=(1-\frac{2M}{r})^{1/2}$ and $\eta_{ab}={\rm diag}(-1,1,1,1)$.
The Riemann tensor is conveniently expressed as
\begin{equation}
R^{\m\n\s\t} = -\frac{M}{r^3} \left[ g^{\m\s}g^{\n\t} -g^{\m\t}g^{\n\s}
+3 U^{\m\n}_{01} U^{\s\t}_{01} -3 U^{\m\n}_{23} U^{\s\t}_{23}  \right]
\label{Riemann}	\end{equation}
using the bivectors
\begin{equation}
U^{\m\n}_{ab}=e_a{}^{\m}e_b{}^\n -e_a{}^\n e_b{}^\m\ .
\label{bivect}	\end{equation}

In the Schwarzschild background, the leading order equations
(\ref{new3}) may be written as:
\begin{equation}
\left[ k^2\, \d^a{}_b + \z\, X^a{}_b \right]\,a^b =0
\label{new3b}
\end{equation}
where $\z=2\b\l^6{M^2\over r^6}$ and using $k_a=e_a{}^\m\,k_\m$
\begin{equation}
X^a{}_b=
\left( \begin{array}{cccc}
    -k_1^{\,2}A+k^2m^2 & -k_0k_1A & k_0k_2\,k^2&
k_0k_3\,k^2\\
    k_0k_1A& k_0^{\,2}A+k^2m^2 & -k_1k_2\,k^2 &
-k_1k_3\,k^2 \\
    -k_0k_2\,k^2  & -k_1k_2\,k^2 & k_3^{\,2}B-k^2l^2 &
         -k_2k_3 B \\
    -k_0k_3\,k^2  & -k_1k_3\,k^2 & -k_2k_3 B &
         k_2^{\,2}B-k^2l^2
	\end{array}\right)
\label{mat3}
\end{equation}
where we have defined $l^2\equiv k_0^{\,2}-k_1^{\,2}$, $m^2\equiv
k_2^{\,2}+k_3^{\,2}$, $A\equiv (4l^2+5m^2)$, and $B \equiv (5l^2+4m^2)$.
Note that $k^2=-l^2+m^2$.
To have a non-trivial solution of
(\ref{new3b}), the determinant of the matrix in square brackets
must vanish.
\begin{equation}
k^2 [k^2(1+\z k^2)] [k^2 +4\z l^4+4\z l^2 m^2+\z m^4]
[k^2 +4\z m^4+4\z m^2l^2 +\z l^4]=0
\label{lcthird} \end{equation}
Here the four factors actually correspond to the eigenvalues of the
matrix, and the polarization associated with a given light-cone
condition will be given by the corresponding
eigenvector. To simplify the analysis, consider photon
trajectories in the plane $\theta=\pi/2$, which
corresponds to $k_2=0$ (or $k_\th=0$), in which case $m^2=k_3^{\,2}$.
\begin{enumerate}
\item
For the eigenvalue, $k^2$, the polarization eigenvector is
$a^b=(-k_0,k_1,0,k_3)=k^b$. This is, of course, the expected
unphysical polarization which associated with the leading order
gauge invariance, $a_\m\rightarrow a_\m+k_\m$ --- \ie since the
gauge condition (\ref{lead2}) was not used in the $O(\b)$ term
in eq.~(\ref{new3}), one must have $X^a{}_bk^b=0$.
\item
The polarization corresponding to the second eigenvalue,
$k^2(1+\z k^2)$, is \\ $a^b=(-k_0k_3,k_1k_3,0,{k_0}^2-{k_1}^2)
=k_3k^b-(0,0,0,k^2)$. Now the vanishing of the eigenvalue
yields $k^2=0$ or $k^2=-1/\z$.
The latter result is nonperturbative in the coupling,
and so the only  relevant result is the
original light-cone $k^2=0$. Note that in this case
the polarization is degenerate with that in case (i),
and so it is also unphysical.
Further note that $k\cdot a=0$ is satisfied here
independent of whether or not one imposes $k^2=0$.
\item
For the third eigenvalue, one finds a generalized light-cone
condition:
$k^2 +\z (4l^4+4 l^2 m^2+ m^4)=0$. The corresponding
polarization is $a^b=(-k_1,k_0,0,0)$, which also satisfies
the Lorentz gauge condition. Thus this case corresponds to
photons with a radial polarization.
\item
The final eigenvalue yields
$k^2 +\z(4 m^4+4 m^2l^2 + l^4)=0$ for a polarization
$a^b=(0,0,1,0)$ along the $z$-axis
--- \ie orthogonal to the $\th=\pi/2$ plane.
Note that the Lorentz gauge condition is also satisfied here.
\end{enumerate}
To leading order, one has $k^2=O(\z)$ in either of the last two cases.
Hence if one applies $k^2=0$ in the $O(\z)$ contributions, one has
$l^2=m^2$ and so (iii) and (iv) both produce the same
perturbed light-cone condition
$k^2+9\z\, l^4=0$. Thus to the order of accuracy which we are
calculating there is no gravitational birefringence here --- \ie
all physical polarizations will follow the same trajectories.
One may also note that for radial motion ($k_3=0$), one has $l^2=m^2=0$
in the $O(\z)$ contribution,
and this light-cone reduces to the usual $k^2=0$.

Now we calculate the modifications to the deflection angle for
photon trajectories which begin at $y\rightarrow-\infty$ and approach
the central mass parallel to the $y$-axis in the $x$-$y$
plane with an impact parameter
$b$ \cite{MTW,Weinberg}.
In the Schwarzschild background using standard coordinates,
the generalized light-cone condition is
\[
-\left( 1-\frac{2M}{r} \right) (1-9\z l^2) k^t k^t + \left(
1-\frac{2M}{r} \right)^{-1} (1-9\z l^2) k^r k^r +r^2 k^\th k^\th +r^2
\sin^2 \th \, k^\p k^\p=0\ \ .
\]
Working perturbatively in $\z$, it is sufficient to use the classical value
for $l^2= k_0 k_0 -k_1 k_1=E^2 b^2 /r^2$ above, where
$E$ is the energy of the photon and $b$, the impact parameter \cite{MTW}.
 Following Ref.\cite{loop2}, the simplest way to determine the
deflection angle from the light-cone
condition is to consider the wave vector $k_\m$ as  a null vector in an
effective metric
\begin{equation}
ds^2=-B(r) dt^2 +A(r) dr^2 +r^2 d\th^2 +r^2 \sin^2 \th \, d\p^2
\label{eff metric}
\end{equation}
with
\begin{eqnarray*}
	A(r) &=& \left( 1-\frac{2M}{r} \right)^{-1} \left( 1- \frac{9\z
E^2 b^2}{r^2} \right) \\
	B(r) &=& \left( 1-\frac{2M}{r} \right) \left( 1- \frac{9\z
E^2 b^2}{r^2} \right)\ \ .
\end{eqnarray*}
The  deflection angle is then given by \cite{Weinberg}
\begin{equation}
\D\p+\pi= 2 \int_{r_0}^{\infty} \frac{dr}{r} \left[
\frac{A(r)}{\frac{r^2}{r_0^2}\frac{B(r_0)}{B(r)}-1}\right]^{1/2}
\label{angle} \end{equation}
 where $r_0$ is the distance of closest approach (at which $k^r$ changes sign).
For the unperturbed
Schwarzschild metric, eq.~(\ref{angle}) yields $\D\p=4M/r_0$ \cite{Weinberg}.
Expanding $A(r)$ and $B(r)$ in
power  of $\z$ and keeping only the linear terms, one obtains an
integral for $\d\D\p$
\[
	\d \D \p = \int_{r_0}^\infty \frac{dr}{r} \left[ \frac{\d
A(r)}{(\frac{r^2}{r_0^2}-1)^{1/2}} -\frac{r^2}{r_0^2} \frac{\d B(r_0)
-\d B(r)} {(\frac{r^2}{r_0^2}-1)^{3/2}} \right]\ \ .
\]
Inserting $\d A(r) =\d B(r) =-9\z E^2 b^2/r^6$, one obtains for
both polarizations
\begin{equation}
\d \D \p =\frac{2205\pi}{128} \frac{\b \l^6 M^2 E^2}{r_0^6}\ \ .
\label{dphiI3}  \end{equation}
Note that both this result and $\D\p=4M/r_0$ are leading order
expressions, which are corrected by terms which are higher order
in $M/r_0$.

A second interaction which produces similar dispersive results is
\begin{equation}
I_4 =-\frac{\b\l^6}{4} \int d^4x \, \sqrt{-g} R^{\a\k\b\l}R_{\a\m\b\n} \na_\k
F_{\l\r} \na^\m F^{\n\r} \ \ .
\label{fourr}
\end{equation}
In this case, the equation of motion for the
electromagnetic field becomes
\begin{equation}
\na_\m F^{\m\n} -\b\l^6 \,\na_\r\na_\m\left(R_{\a\k\b\l}
R^{\a\m\b[\r|} \na^\k F^{\l|\n]}\right) =0
\end{equation}
where the square brackets indicate that the expression is
antisymmetrized in $\r$ and $\n$ with a factor of $1/2$.
Inserting the geometric ansatz (\ref{ans}) and applying the gauge
constraint (\ref{lead2}) produces the leading order equation
\begin{equation}
k^2 a^\n +\b\l^6 R_{\a\k\b\l} R^{\a\m\b\r} k_\r k_\m k^\k k^{[\l} a^{\n]} =0.
\label{new4}
\end{equation}
One may now follow the procedure used above to determine the
modified light-cone condition for photons in a Schwarzschild background.
A simpler approach for this specific case yields a general light-cone
condition --- namely, contract eq.~(\ref{new4}) with $a_\n$.
Upon applying the gauge condition (\ref{lead2})  and
extracting a factor of $a^2$, one obtains
\begin{equation}
k^2 +\frac{\b\l^6}{2} R_{\a\m\b\r} R^{\a\k\b\l} k_\l k_\k k^\m k^\r =0
\label{lcfourth} \end{equation}
Thus we have a general light-cone condition which describes all
polarizations, and
so there will be no gravitational birefringence in any background.
(Note that the unphysical polarizations do not satisfy $k^2=0$ in this
case, because in arriving at eq.~(\ref{lcfourth}) we have used the
gauge constraint in the $O(\b)$ terms).
In the Schwarzschild background, this light-cone becomes
\begin{equation}
k^2 + {\z\over 4}\left[ 9l^4+9m^4-3k^4  \right]
\label{lcfrty}
\end{equation}
where as above, we use $\z=\frac{2\b\l^6 M^2}{ r^6}$,
$l^2= k_0^{\,2}-k_1^{\,2}$, and $m^2=k_2^{\,2}+k_3^{\,2}$.
To leading order, $k^2={\cal O}(\b)$ and $m^2=l^2+{\cal O}(\b)$ and so
eq.~(\ref{lcfrty}) reduces to $k^2+\frac{9}{2}\z l^4$.
Hence up to a factor of two, we have recovered precisely the
same dispersive light-cone as in the analysis of the interaction $I_3$
(for the physical polarizations).
The modification to the deflection of light is therefore one half
the angle obtained in  eq.~(\ref{dphiI3}).

A final eight-derivative interaction which produces dispersive light
propagation was found by extending eq.~(\ref{second}) by introducing
extra background derivatives, rather than an extra curvature
tensor
\begin{equation}
I_5=-\frac{ \b \l^6}{4} \int d^4x \, \sqrt{-g}
\na_{(\r} \na_{\s)} R_{\a\b\m\n} \na^\r F^{\a\b} \na^\s F^{\m\n}
\label{fivve}
\end{equation}
where $\na_{(\r} \na_{\s)}=1/2(\na_\r\na_\s+\na_\s\na_\r)$.
After adding $I_5$ to the Maxwell action (\ref{max}),
the equation of motion for the electromagnetic field becomes
\[
	\na_\m F^{\m\n} -\b \l^6\,\na_\m\na_\s\left(\na^{(\s}\na^{\r)}
        R^{\a\b\m\n} \na_\r F_{\a\b}\right) =0\ \ \ .
\]
We then insert the geometric optics ansatz (\ref{ans}) to obtain:
\begin{equation}
	k^2 a^\n +2\b \l^6  \na^\s \na^\r R^{\a\b\m\n} k_\s k_\r k_\m
	k_\a a_\b =0
\label{I5approx} \end{equation}
To calculate the light-cone conditions in the Schwarzschild background,
we followed the same method that was used in the analysis of
$I_3$ above. The leading order equations may be written
\begin{equation}
\left[ k^2 \d^a{}_b +\h \widehat{X}^a{}_b \right] a^b=0
\end{equation}
where $\h=60\b\l^6 M/r^5$. The matrix $\widehat{X}^a{}_b$ is a lengthy
expression involving $k_a$, which we will not display explicitly
here. As before, the
modified light-cone conditions are given by the vanishing of
the eigenvalues of the matrix in square brackets,
and the corresponding eigenvectors give the associated polarizations.
If we consider photon trajectories in the plane $\th=\pi/2$ and
we apply $k^2={\cal O}(\b)$ in $\widehat{X}^a{}_b$, we find:
\begin{enumerate}
\item
$k^2=0$ for the two unphysical polarizations
\item
$k^2+\h Cl^2=0$ for the radial polarization
\item
$k^2-\h C l^2=0$ for the polarization orthogonal to the plane of motion.
\end{enumerate}
where
\begin{equation}
C=\left( 1-\frac{3M}{r} \right) k_0^{\,2}
-\left(7-\frac{15M}{r} \right) k_1^{\,2}\ \ .
\end{equation}
Since the light-cones for the radial and theta polarizations differ,
this last case provides an example of gravitational
birefringence. Calculating the deflection angle as above, we
find an energy-dependent contribution
\begin{equation}
\d\D\p=\pm504 \frac{\b\l^6 M E^2}{r_0^5}
\label{dDphi} \end{equation}
where the plus sign corresponds to the radial polarization and
the minus sign, to the theta polarization. Note that this result
is one order lower in the $M/r_0$ expansion than the previous
result (\ref{dphiI3}). This reduction occurs since the present
interaction (\ref{fivve}) involves a single Riemann tensor, while
the previous interactions have two curvature tensors.

\section{Discussion}

In this paper, we have found some explicit field theory interactions
that produce dispersive photon propagation, in
the context of an effective field theory where the
Maxwell action is modified by higher-derivative terms.
Such dispersion was not observed in earlier studies simply
because the effective actions considered previously did not
include sufficiently high numbers of derivatives. The
final case also provides a new example of gravitational birefringence.
One may ask whether there will be other eight-derivative
interactions which will produce dispersion, and clearly the answer
is yes. The three interactions that we have considered though
are representatives of three classes of interactions, which produce
the same leading order equations in the geometric optics approximation.
It is not hard to verify that eq.'s (\ref{new3}), (\ref{lcfourth})
and (\ref{I5approx}) are unique. For instance, there is only
a single way to contract four wave-vectors and one polarization with a
double derivative of the background curvature tensor, and this
is the combination appearing in eq.~(\ref{I5approx}). Thus,
\begin{equation}
I'_5=-\frac{ \b \l^6}{2} \int d^4x \, \sqrt{-g}
\na_{(\r} \na_{\s)} R_{\a\b\m\n} \na^\r F^{\s\a} \na^\b F^{\m\n}
\end{equation}
leads to precisely eq.~(\ref{I5approx}) as the leading order equations
of motion. The two interactions, $I_5$ and $I'_5$, differ by
total derivatives, and also terms which do not contribute to these
leading order contributions (\ie they do not contribute to
the dispersion).

In eq.'s (\ref{dphiI3}) and (\ref{dDphi}), we have found contributions
to the deflection angle of the light rays, which depends on the square
of the photon energy. This behavior is the same as that found for string
theory by Mende. One may ask then if interactions of the form
discussed here appear amongst the higher dimension
interactions included in the low-energy effective string action.
There are two alternative approaches to constructing
these low-energy actions. First, it can be determined from the sigma
model beta functions, which define the low energy string
equations of motion\cite{string,sigmacomp}. Unfortunately,
sigma model
calculations involving background metric and gauge fields
have not been carried out to sufficient order to detect terms
of the form suggested here. Alternatively, the higher-derivative
terms in the low-energy action
can be determined by requiring that the resulting field theory
reproduce the string scattering amplitudes to the corresponding
order in $\ap p^2$ \cite{string,scatcomp}, where $p$ here represents
a typical momentum from the scattering process, and
$\ap=\lambda_{st}^2$ is the string scale squared (essentially, this
corresponds to the Planck scale squared).
Interactions of the form (\ref{third}), (\ref{fourr})
or (\ref{fivve}) would contribute to an scattering amplitude
of two photons and two gravitons --- the contribution of $I_5$
to a two photon and one graviton amplitude vanishes on-shell. A
sufficiently detailed study of the low-energy effective
action for heterotic string has been made to detect
terms of the form discussed here\cite{scatcomp}, but
unfortunately, one finds that these terms do not
appear in this action. This suggests that Mende's dispersive effect,
which should be universal to all string theories\cite{Mende},
must be produced by an interaction at an even higher order in the $\ap$
expansion (or the expansion in numbers of derivatives) than
considered in the present paper.
So one would expect that the dependence on
the radius of closest approach is even more dramatic than the
$r_0^{-6}$ appearing in eq.~(\ref{dphiI3}). Additional
Riemann tensors would also increase the power of the central
mass appearing in the dispersive contribution to the deflection
angle.

If one considers studies of low-energy string actions, there
is one eight-derivative interaction which is in fact universal
to all string theories\cite{super}
\begin{equation}
I_6={\z(3)\ap^3\over512\pi}\int d^Dx\,\sqrt{-g}\left(2 R_{\a\b\r\s}
R_\m{}^{\b\r}{}_\n R^{\a\g\h\m} R^\n{}_{\g\h}{}^\s
+R_{\a\b\r\s}R_{\m\n}{}^{\r\s}R^{\a\g\h\m}R^\n{}_{\g\h}{}^\b\right)
\end{equation}
where $\z(s)$ is the Riemann zeta function.
We have also indicated that this effective action
is in $D$ dimensions, since typically string theories are
constructed for $D>4$. If the spacetime is then compactified
down to four dimensions via a Kaluza-Klein ansatz\cite{kakl}, then
new vector particles will appear in the effective theory arising
from off-diagonal components of the metric, which mix the four-dimensional
spacetime with the compact directions (\eg $g_{\m5}\simeq A_\m$).
The $D$-dimensional Einstein action provides the standard
Maxwell action (\ref{max}) for these vectors upon compactification.
Similarly one finds that upon compactification the above interaction
yields interactions of the form of eq.'s (\ref{third})
and (\ref{fourr}) (\eg using $R_{5\a\b\m}\simeq-{1\over2}
\nabla_\a F_{\b\m}+\ldots$). Therefore the
above string interaction produces dispersive propagation as described
in our present analysis for these
Kaluza-Klein vector fields. The latter, of course, correspond
to particular modes in the string spectrum.

Finally, we consider the magnitude of the deflection angles that
we have calculated. Ultimately, we expect that this dispersion
would only be observable in very exotic circumstances, but
to begin let us evaluate eq.~(\ref{dDphi})
with solar parameters for which the leading order
deflection angle of general relativity is
$\D\p=4M/r_0=1''.75$ \cite{MTW}. The length scale $\l$ is
the microphysical scale associated with the processes
that induce our effective interaction. Here,
we will choose the
interaction scale to correspond the Compton wavelength
of the electron (\ie $\lambda=\l_e\simeq2.4\times 10^{-12}$ m)
as it would be if eq.~(\ref{fivve}) arose
as a higher order term in the derivative expansion of the
one-loop effective action for QED --- clearly the
effect will be more suppressed if we chose a shorter length scale,
\eg the Planck scale in a string effective action.
In this case, it is natural to choose the dimensionless
coupling constant to be of the order
of the fine structure constant (\ie $\beta\simeq\a$). With these
choices, the dispersive deflection angle for a radially polarized
photon grazing over the limb of the sun (\ie $r_0\simeq7\times10^8$ m)
is given by
\begin{equation}
\frac{\d\D\p}{\D\p} \simeq  {10^{-89} \over\lambda_{ph}^2}
\end{equation}
where $\lambda_{ph}$ is the wavelength of the photon measured in
angstroms. So the visible spectrum ranging from four- to seven-thousand
angstroms would be spread over an angle of about $6\times10^{-98}$
arcseconds.  Clearly as such, gravity's rainbow would be
unobservable.

Now we also wish to consider situations in which the dispersion
would become more pronounced. If we are consider eq. (\ref{dphiI3})
or (\ref{dDphi}) with $\l$ and $\b$ fixed as above for the QED
(\ie $\l=\l_e$ and $\b\simeq\a$), there are three options:
increase the photon energy, decrease the radius of closest approach
or increase the central mass. With any of these options, we are
limited by the approximations entering into our calculations.
The deflection of much higher energy photons is certainly
greater, but one must remember that the
applicability of the effective action limited to photon wavelengths
greater than the interaction scale $\lambda$, which we are here considering to
be the Compton wavelength of the electron. Thus one could only consider
photons up to the X-ray portion of the spectrum.
The deflection is also increased with a reduction in
the radius of closest approach $r_0$. This radius would be minimized
by considering a black hole for which one might achieve $r_0\simeq M$.
Such a scenario, though, runs
into conflict with another approximation made in our scattering angle
calculations, namely $M/r_0\ll 1$. In principle,
one could carry out those calculations in more detail if one wished
to consider $M/r_0\simeq1$. With this choice then, one would
actually want to decrease, rather than increase, the mass, $M$.
Here the limitation is the validity of the
geometric optics approximation, which requires that the
photon wavelength be much smaller than the radius of curvature
of the spacetime geometry. In a Schwarzschild geometry then,
one demands that $\lambda_{ph}^2<r_0^3/M\simeq M^2$. Thus
at least $M$ must be greater than $\l_e$, which was a lower
bound on the photon wavelength. Certainly, one could
imagine then that dramatic dispersion
would be produced for X-rays by a black hole of $M\simeq 10^{18} {\rm g}$,
for which the gravitational radius would be of the order of
the Compton wavelength of electron. It seems, though, that such
an object (with $M\simeq 10^{-15}M_\odot$) and the dispersed
X-rays are unlikely to be observed.
It may also be interesting though to consider photon propagation
beyond the geometric optics approximation. It may be that
effective interactions of lower dimension than considered in
section 3 could produce dispersion in situations with
large and rapidly varying curvatures, as could possibly be
created by gravitational collapse. In conclusion, while the
dispersive photon propagation appearing in the present analysis
in principle presents a violation of the equivalence principle, it
appears to be beyond the practical limits of observations.

\section*{Acknowledgments}
This research was supported by NSERC of Canada and Fonds FCAR du
Qu\'ebec. We would like to thank Cliff Burgess,
Paul Mende and Joe Polchinski for useful conversations.


\begin{thebibliography}{99}
\bibitem
{foot1}
Berends and Gastmans\cite{Berends} reported to find energy-dependent
light deflection within the context of the one-loop effective action
from QED, but this was a erroneous result, as described by
Drummonds and Hathrell\cite{loop2}.
\end{thebibliography}
\end{document}